\documentclass[twocolumn]{aastex62}

\newcommand{\vcn}{V$^*$ CN Cha }

\newcommand{\diff}{\mathrm{d}}
\newcommand{\gcheck}{\textcolor{green}{\checkmark}}
\newcommand{\exmark}{\textcolor{red}{X}}
\newcommand{\unsure}{\textcolor{blue}{?}}

\graphicspath{{./}{figures/}}

\received{XXX}
\revised{XXX}
\accepted{\today}
\submitjournal{AJ}

\shorttitle{A Mystery in Chamaeleon}
\shortauthors{Lancaster et al.}

\begin{document}

\title{A Mystery in Chamaeleon: Serendipitous Discovery of a Galactic Symbiotic Nova}

\correspondingauthor{Lachlan Lancaster}
\email{lachlanl@princeton.edu}
\author[0000-0002-0041-4356]{Lachlan Lancaster}
\affiliation{Department of Astrophysical Sciences, Princeton University, 4 Ivy Lane, 08544, Princeton, NJ, USA}

\author{Jenny E. Greene}
\affiliation{Department of Astrophysical Sciences, Princeton University, 4 Ivy Lane, 08544, Princeton, NJ, USA}

\author[0000-0001-5082-9536]{Yuan-Sen Ting}
\altaffiliation{Hubble Fellow}
\altaffiliation{Carnegie-Princeton Fellow}
\affiliation{Institute for Advanced Study, Princeton, NJ 08540, USA}
\affiliation{Department of Astrophysical Sciences, Princeton University, 4 Ivy Lane, 08544, Princeton, NJ, USA}
\affiliation{Observatories of the Carnegie Institution of Washington, 813 Santa Barbara Street, Pasadena, CA 91101, USA}
\affiliation{Research School of Astronomy and Astrophysics, Mount Stromlo Observatory, Cotter Road, Weston Creek, ACT 2611, Australia}

\author[0000-0003-2644-135X]{Sergey E. Koposov}
\affiliation{McWilliams Center for Cosmology, Carnegie Mellon University, 5000 Forbes Avenue, 15213, Pittsburgh, PA, USA}
\affiliation{Institute of Astronomy, University of Cambridge, Madingley Road, Cambridge, CB3 0HA, UK}
\affiliation{Kavli
 Institute for Cosmology, University of Cambridge, Madingley Road, Cambridge CB3 0HA, UK}
\author[0000-0003-2595-9114]{Benjamin J. S. Pope}
\altaffiliation{NASA Sagan Fellow}
\affiliation{Center for Cosmology and Particle Physics, Department of Physics, New York University, 726 Broadway, New York, NY 10003, USA}
\affiliation{Center for Data Science, New York University, 60 Fifth Ave, New York, NY 10011, USA}

\author[0000-0002-1691-8217]{Rachael L. Beaton}
\altaffiliation{Hubble Fellow}
\altaffiliation{Carnegie-Princeton Fellow}
\affiliation{Department of Astrophysical Sciences, Princeton University, 4 Ivy Lane, 08544, Princeton, NJ, USA}
\affiliation{Observatories of the Carnegie Institution of Washington, 813 Santa Barbara Street, Pasadena, CA 91101, USA}

\begin{abstract}

We present the serendipitous discovery of a low luminosity nova occurring 
in a symbiotic binary star system in the Milky Way. We lay out the extensive 
archival data alongside new follow-up observations related to the stellar 
object \vcn in the constellation of Chamaeleon. The object had long period 
($\sim\! 250\,$day), high amplitude ($\sim\! 3\,$mag) optical variability 
in its recent past, preceding an increase in optical brightness by 
$\sim\! 8\,$magnitudes and a persistence at this luminosity for about 3 
years, followed by a period of $\sim\! 1.4\,{\rm mag}\,{\rm yr}^{-1}$ 
dimming. The object's current optical luminosity seems to be dominated by 
H$\alpha$ emission, which also exhibits blue-shifted absorption (a 
P-Cygni-like profile). After consideration of a number of theories to 
explain these myriad observations, we determine that \vcn is most likely 
a symbiotic (an evolved star-white dwarf binary) system which has undergone 
a long-duration, low luminosity, nova. Interpreted in this way, the outburst 
in \vcn is among the lowest luminosity novae ever observed.

\end{abstract}



\section{Introduction}
\label{sec:intro}

\begin{center}
    \textit{``One of the advantages to being disorganized is 
    that one is always having surprising discoveries."}\\
     - A.A. Milne \citep{winnie_the_pooh}
\end{center}
\vspace{-0.1in}

The array of human-collected astronomical data is vast. Varied in both 
structure and content, there is no single archive that gathers all such
data in to a single, easily accessible place. Comparison between historical 
and contemporary datasets can reveal many previously-overlooked astrophysical
phenomena. In this paper, we present a dramatic outburst from the star 
V$^*$ CN Cha, whose enigmatic nature was only revealed through the broad 
temporal and spectral coverage available across astronomical archives.

\begin{figure*}
\includegraphics{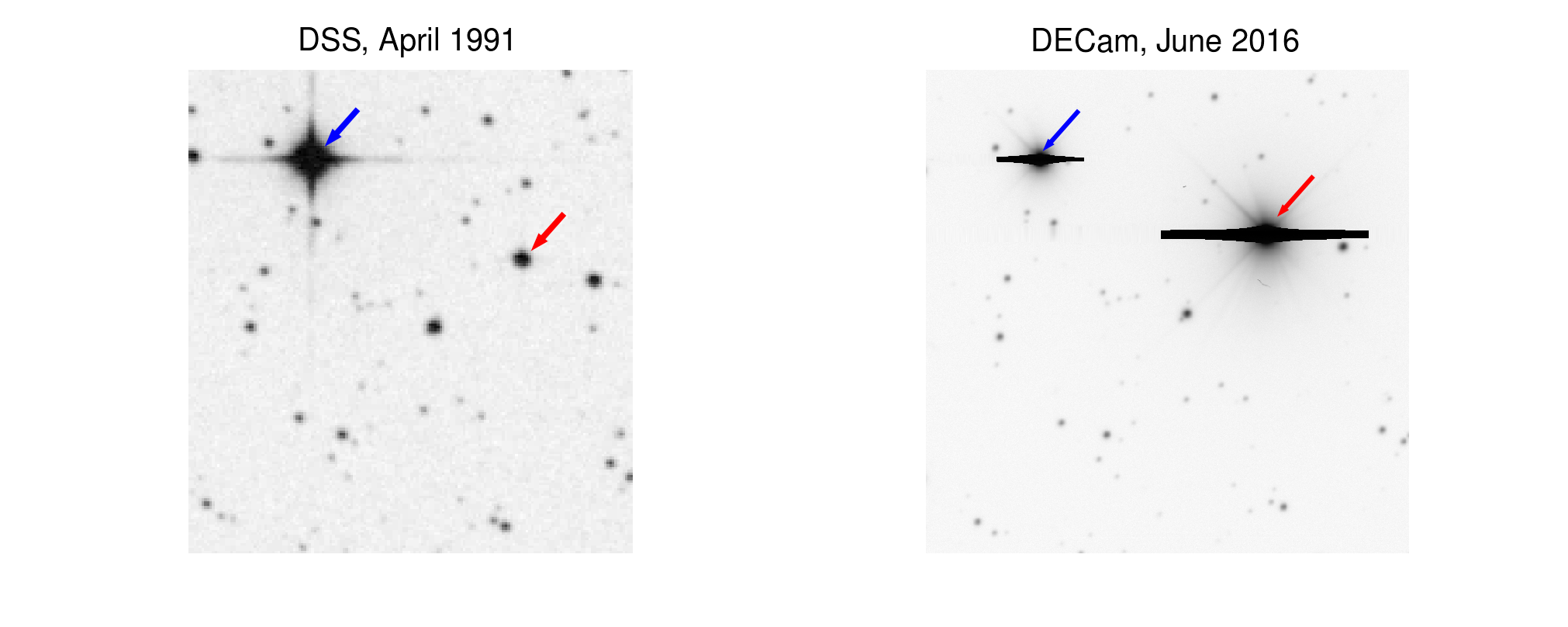}
\caption{We show a comparison of two images of \vcn, which is indicated by 
a red arrow in both panels. The left panel shows the DSS $R$ band observation 
taken in April of 1991 and the right shows the DECam $r$ band observation 
taken in June of 2016. Despite the fact that the the images are at different 
contrast levels and resolutions, and they are taken in slightly different 
bands, it is clear from comparison to the nearby star CD-79 452 (indicated 
with a blue arrow) that \vcn has increased in brightness considerably. }
\label{fig:image_comp}
\end{figure*}

On a recent observing run, we chose bright and distant 
objects from the \textit{Gaia} DR2 data 
\citep{GAIA_DR2_mission,GAIA_DR2_contents} for spectroscopic follow-up. 
One such candidate was the star \vcn in the constellation of Chamaeleon. 
As we will outline below, the observations that followed were 
quite intriguing. Yet the archival data which had already been 
gathered on this object were even more fascinating, including variations 
in brightness over a range of ten magnitudes that occurred on a 
variety of time scales. This variation is exemplified in 
Figure~\ref{fig:image_comp}, which shows observations of \vcn 
separated by 25 years.

In Section \ref{sec:data} we present and summarize the set of 
archival data pertaining to this object. We then 
present the optical spectrum that we observed for \vcn in Section 
\ref{sec:our_obs}. After having presented the data we provide 
more detailed analysis of various parts of the data in Section 
\ref{sec:deduction}. Using the net sum of the observational data we 
then discuss the possible theories that could explain them in 
Section \ref{sec:theories}. Among these theories, we determine that 
it is likely that \vcn is a symbiotic nova. In Section~\ref{sec:discussion} 
we briefly discuss the implications of the interpretation of \vcn as a 
symbiotic nova on the study of novae. Finally, we discuss possible 
follow-up observations that could be taken of this object and 
conclude in Section \ref{sec:conclusion}.

\section{Archival Data}
\label{sec:data}

We present the various archival data that we have gathered on this object. 
We begin with a summary of the data as a whole. This is followed by a detailed 
description of each constituent data set, beginning with the \textit{Gaia} 
data on the object, which first motivated our investigation. We then present 
the numerous archival photometric observations, roughly in chronological order, 
so as to give the reader a full explanation of the data underlying the 
photometric variability.

\subsection{Archival Summary}
\label{subsec:phot_summary}

\vcn was first identified as a Mira variable in 1963 and was observed 
by multiple surveys throughout the 20th century which took measurements 
consistent with this initial conclusion. A visual summary of the 
photometric data since 2000 is given in 
Figure~\ref{fig:summary_plot}. We gather multi-epoch photometry spanning 
nearly 20 years ranging from the near UV to the far infrared. The 
information summarized in Figure~\ref{fig:summary_plot} indicates that 
\vcn experienced an outburst event at some point in 2013. We find that the 
pre-outburst SED is best described by the sum of two blackbodies, one cool
($2000\,$K) bright component, and one hot ($10000\,$K) dim component.
Observing the 
SEDs from before and after this event, it is clear that the object became 
much brighter in almost all bands for which we can make a reasonable 
comparison. This brightness difference is most extreme in the UV where an 
observed increase in flux by a factor of $10,000$ is seen between UVOT and 
SkyMapper measurements. It should be noted that these SkyMapper measurements 
in the UV were taken very soon after the outburst event, in March of 2014, so 
it is plausible to assume that the SED now is somewhat different than it was 
then, especially given the dimming that was observed later in the ASAS-SN 
lightcurve.

Finally, we show the TESS lightcurve in Figure~\ref{fig:tess_lc}, which we 
classify as `archival' despite the fact that it was taken after our 
own observation. The main conclusions from the TESS data are that 
(1) they indicate that the star is still steadily dimming at a rate 
consistent with the ASAS-SN observations and (2) they suggest the 
star has short-term variability on the timescale of several hours.

\subsection{Gaia}
\label{subsec:gaia}

The European Space Agency's astrometric space mission \textit{Gaia}, 
observed \vcn during its first window of observation used 
in the mission's second data release (DR2) 
\citep[DR2][]{GAIA_DR2_mission,GAIA_DR2_contents}. This observation window 
was from July 24th, 2014 to May 23rd, 2016. The mission observed 
the star to be peculiar in a number of ways that alerted the authors 
that the star may be interesting. The key observations 
that piqued our interest were:
\begin{itemize}
    \item[1.] The photometric observations gave $G=7.41$, 
    $G_{\rm BP}=7.72$, $G_{\rm RP} = 6.98$ mag, making the star
    relatively bright.
    \item[2.] The star was assigned a parallax of 
    \begin{equation}
        \varpi = 0.3142 \pm 0.0249 \, {\rm mas}
    \end{equation}
    putting the star's distance at $3.18^{+0.27}_{-0.23}\,$kpc, 
    implying an absolute $G$ band magnitude of $-5.1$ mag, making 
    it unusually luminous.
    \item[3.] The astrometric pipeline assigned zero astrometric 
    excess noise (AEN) to this object, which indicates that the 
    astrometry does not show obvious problems. The Renormalized 
    Unit Weight Error (RUWE) is $1.19\,$.
    \item[4.] The star was not classified as variable.
\end{itemize}

The first three of the above items were enough to identify this object 
as one of interest for follow-up.

\subsection{Identification}
\label{subsec:identify}

\vcn was first identified as a Mira variable star in 1963 as reported 
by \cite{Hoffmeister63}. At that time the star was identified to vary 
between magnitude 15 and 
17\footnote{\url{www.sai.msu.su/gcvs/cgi-bin/search.cgi?search=CN+Cha}}, 
in the Johnson $I$ band.

\subsection{IRAS}
\label{subsec:IRAS}

The star was later identified by the Infrared Astronomical Satellite (IRAS) 
as a strong point source in all four bands (12, 25, 60, and 100 $\mu$m) as 
well as being given an 82\% probability of being variable based on analysis 
of the 12 and 25 micron flux densities and their uncertainties \citep{IRAS}.

\subsection{Hipparcos}
\label{subsec:Hipparcos}

Despite being very bright in the optical at the current epoch, it was not 
recorded as a source in any of the catalogues created from observations 
performed by the Hipparcos satellite \citep{HippTycho,Tycho2}. The 
completeness of the Tycho-2 catalog is 99\% for stars brighter 
than 11 magnitude in the $V$ band \citep{Tycho2}, suggesting that \vcn was 
likely dimmer than 11$^{\rm th}$ magnitude in the $V$ band during 
Hipparcos's observations from 1989 to 1993.

\subsection{Digitized Sky Survey}

The Digitized Sky Survey (DSS) consists of a series of photometric plate 
observations done on the Oschin and UK Schmidt Telescopes between 1983 and 
2006 \citep{DSS_POSSII96,DSS_GSC90}. \vcn was observed by the UK 
Schmidt Telescope using the \texttt{RG610} filter (approximating the 
Johson-Cousins $R$ band) on April 12th, 1991. We show this observation in the 
left hand panel of Figure \ref{fig:image_comp}. The star was reported to have 
magnitude \texttt{RG610}=13.26 mag in this band by the DSS \citep{DSS_GSC90}.


The star was also observed by the DSS in January of 1976 in a 
\texttt{GG395} filter at a bluer wavelength and in March of 1996 in a 
\texttt{RG715} filter at a redder wavelength. While these exposures 
are available online\footnote{\url{https://archive.stsci.edu/cgi-bin/dss_form}}, 
we could not determine zero-point fluxes for stars in these bands or 
stars within the field with reference magnitudes in these bands, so we 
do not report any photometry for these images.

\subsection{Deep Near Infrared Survey}
\label{subsec:DENIS}

The Deep Near Infrared Survey (DENIS) was a near infrared survey operated 
out of the 1-meter European Southern Observatory (ESO) telescope at La 
Silla, Chile \citep{DENIS}. DENIS also gathered multiple observations of 
355 million objects in each of its three bands: $I$ ($0.8\, \mu$m), 
$J$ ($1.2\, \mu$m), and $K_s$ ($2.1\, \mu$m)\citep{DENIS_phot}. The DENIS 
survey observed \vcn on three occasions from early 1999 to January of 2000
\citep{DENIS_DR3}. However all observations in the $J$ and $K_s$ bands were 
brighter than the saturation limits of the DENIS survey, as was the 
second observation in the $I$ band \citep{DENIS}. For this reason we only 
use the two DENIS observations in the $I$ band that were unsaturated. 
The range of flux that these observations cover is indicated in 
Figure~\ref{fig:summary_plot}.

\subsection{Two-Micron All Sky Survey}
\label{subsec:2MASS}

The Two-Micron All Sky Survey (2MASS) was a full-sky survey in the $J$, 
$H$ ($1.65\, \mu$m), and $K_s$ bands operated on two dedicated 1.3 meter 
telescopes at Mount Hopkins, Arizona and Cerro Tololo, Chile between June 
of 1997 and February of 2001. The survey was estimated to be complete at 
signal-to-noise greater than 10 to approximately 15th 
magnitude in the $H$ band. \vcn was observed by 2MASS on January 13th, 
2000 \citep{2MASS}. 

\subsection{All Sky Automated Survey}
\label{subsec:ASAS}

The All Sky Automated Survey (ASAS) was designed as one of the first 
major efforts to obtain time-variability information on a large fraction 
of stars in the night sky \citep{ASAS}. 
The ASAS survey obtained observations of \vcn from November 21st, 2000 to 
November 27th, 2009 \citep{ASAS_variables1,ASAS_variables2,ASAS_variables3}. 
The light curve for the star is reported in two separate online catalogues that we were able to find. We use both catalogues, as they 
seem consistent with one another, and discuss the relation between 
them in Appendix~\ref{asas_app}.

The ASAS survey used these observations to identify \vcn as a Mira 
variable star with a period of 260 days. The light curves from the 
original ASAS survey were reanalyzed in 2016 by \cite{ASAS2} with 
more rigorous analysis techniques. In this reanalysis, \vcn has a 
period of 253.24 days.

\subsection{Radial Velocity Experiment}
\label{subsec:RAVE}

The Radial Velocity Experiment (RAVE) obtained a spectrum of \vcn on
February 20th, 2010 \citep{RAVEDR4}. The RAVE team very generously 
provided us with a copy of the spectrum taken in 2010 in advance of 
its planned release in 2020. The star's spectrum shows clearly very 
low $T_{\rm eff}$ with broad TiO absorption features, consistent with 
a cool giant star. Unfortunately, the RAVE pipeline fits seem to have 
hit the boundary of the stellar parameter grids and therefore extracted stellar model parameters are not reliable (RAVE Team, private communication). We leave the analysis 
of this spectrum to future work.

\subsection{AKARI}
\label{subsec:akari}

The AKARI satellite was a Japanese astronomical all-sky survey 
in six infrared bands between $9 \, \mu$m and $200\, \mu$m 
\citep{akari07}. This survey was carried out between May 6th, 2006 
and August 8th, 2007, during which time AKARI observed \vcn and 
reported measurements for its flux in band passes centered at 
$9\, \mu$m and $18\, \mu$m and non-detections in all longer 
wavelength bands at 65, 90, 140, and 160 $\mu$m \citep{akari10_FIS}.

\subsection{Swift/UVOT}
\label{subsec:uvot}

\begin{figure*}
    \centering
    \includegraphics{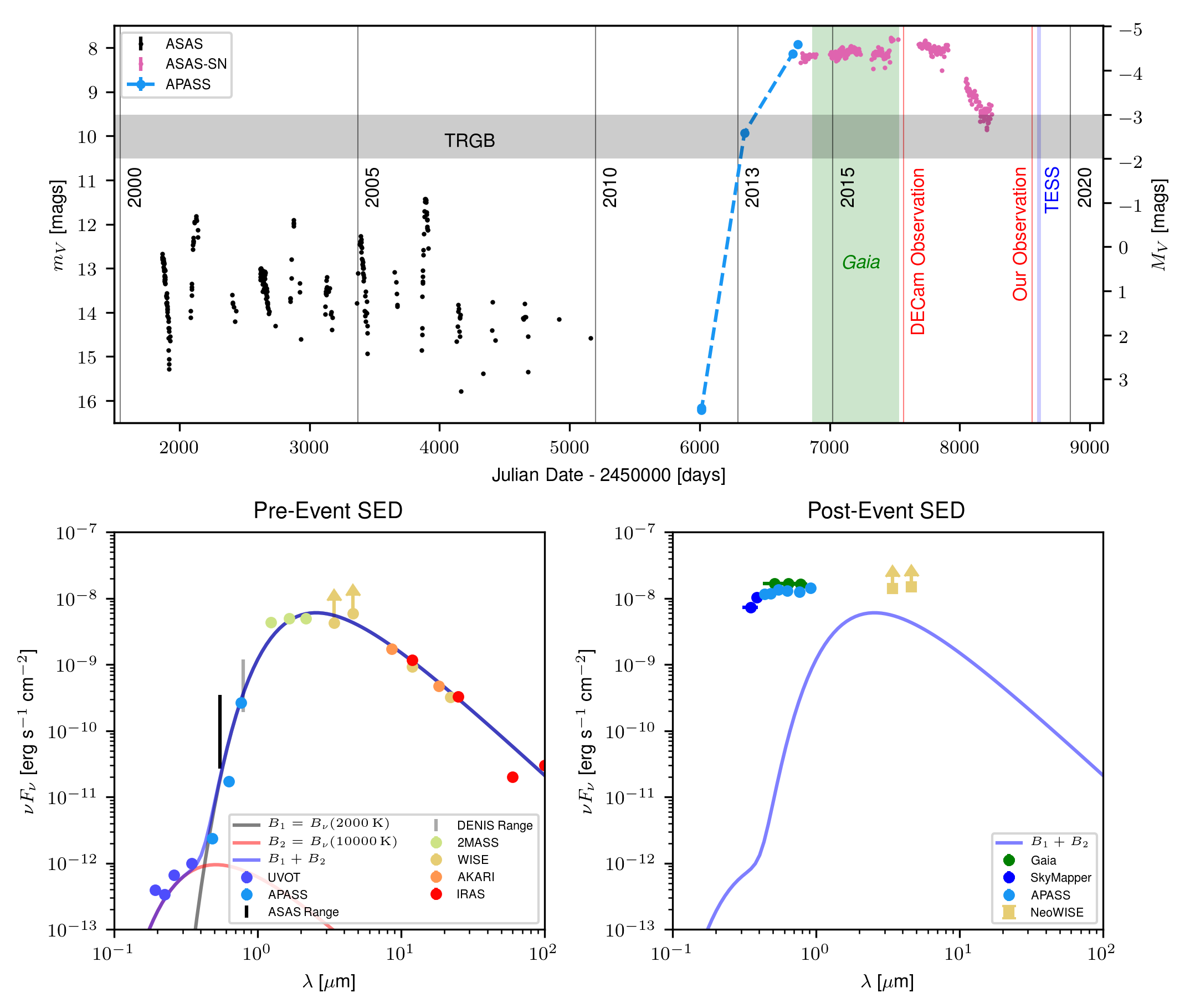}
    \caption{A summary of the photometric data. 
    \textit{Top Panel}: We show a combination of light curves from the ASAS 
    survey (black points), the APASS survey (light-blue points connected by 
    a dashed line) and the ASAS-SN survey (pink points) all in approximations 
    of the Johnson-Cousins $V$-band; we make no correction for marginal 
    differences in the definitions of these bands between surveys. The
    scale for the apparent (absolute, from \textit{Gaia} parallax) magnitude is 
    shown at left (right). We indicate intervals of 5 years (along with 2013, the 
    year we believe the outburst occurred) with vertical grey lines, the DECam 
    and our spectroscopic observations as vertical red lines, and the period 
    over which the \textit{Gaia} mission used data for its DR2 as a green band. 
    The small blue band indicates the TESS observations. We indicate the typical 
    range in absolute magnitude spanned by the tip of the red giant branch 
    (TRGB) as a horizontal grey band.
    \textit{Bottom Left Panel}: We show an SED for the star for photometric data 
    that is determined 
    to have been taken `pre-event.' We compare the data with an SED which 
    is the sum of two blackbody spectra, one with a temperature of $10,000\,$K 
    and one with a temperature of 2000K for reference, the normalization of 
    this spectrum was chosen to marginally follow the observed photometric data. 
    We chose two blackbodies to match the UV excess observed in the data. As the 
    star is observed to vary quite strongly in the `pre-event' phase we 
    attempt to illustrate this variability by indicating the range of variability 
    of the ASAS $V$ band measurements as a black band and the range of the unsaturated
    DENIS measurements as grey bands. Due to the variability the deviation 
    from the blackbody in the  variable region should not be taken as 
    significant. \textit{Bottom Right Panel}: An analog of the bottom left 
    panel but for `post-burst' photometric data. The sum of blackbodies has 
    the same scale as the curve shown in the bottom left panel. The photometry 
    is much more stable in the `post-outburst' phase so we can somewhat safely 
    interpret these points as representing the true SED.}
    \label{fig:summary_plot}
\end{figure*}

The Neil Gehrels \textit{Swift} Observatory is a space observatory 
mission directed by NASA and built for the main purpose of detecting 
Gamma Ray Bursts (GRBs)\footnote{\url{https://swift.gsfc.nasa.gov}}.
The \textit{Swift} UV and Optical Telescope (UVOT) observes the sky 
in four broad bands $UVW2$($1928$\AA), $UVM2$($2246$\AA), 
$UVW1$($2600$\AA), and $U$($3465$\AA) spanning the far to near UV
\citep{UVOT_calib}. Data from UVOT was used to create the 
\textit{Swift}/UVOT Serendipitous Source Catalog (SUVOTSSC) which 
catalogs a number of point sources in these bands 
\citep{Swift_UVOT1,Swift_UVOT2}. \vcn was observed six times by 
UVOT and we display the photometry from January 
of 2008 in Figure \ref{fig:summary_plot}.

\subsection{APASS}
\label{subsec:apass}

The American Association of Variable Star Observers (AAVSO) Photometric 
All-Sky Survey (APASS) is a survey performed primarily in five bands: 
Johnson-Cousins $B$ and $V$ as well as Sloan $g'$, $r'$, and $i'$ 
\citep{apass09,apass_henden14}. With their latest data release (DR10), 
there have been observations in the Sloan $u'$ and $z'$ bands as well as the 
$Y$ band \citep{apass_dr10}.

The epoch photometry from the APASS survey is also available through the 
AAVSO's International Variable Star Index (VSX)
\footnote{\url{https://www.aavso.org/vsx/}}. The VSX data includes 
$g'$, $r'$, and $i'$ values for 5 epochs spanning from March 26th, 2012 
to April 5th 2014. The last three of these 5 epochs additionally have 
photometric measurements in the $B$, $V$, and $z'$ bands. This 
photometric information was essential in constraining the behaviour 
of \vcn in an otherwise unobserved portion of its evolution as shown 
in Figure \ref{fig:summary_plot}, where we can see its rise in 
luminosity from magnitude 16 to magnitude 8.

As we have the most information on the long-term variability of 
\vcn in the $V$-band, in order to have the largest number of epochs 
for comparison we use the APASS $g'$ and $r'$ bands to calculate an 
effective $V$ band magnitude for the two earliest observations where 
there is no directly measured $V$ band magnitude. The calibration we 
use is 
\begin{equation}
    V = g' - 0.52(g'-r') - 0.03 \, \rm{mag}\, ,
\end{equation}
which comes from \cite{Jester05}. It should be noted that this relation 
is derived using relatively well behaved stars, so our use of the relation 
here for such a strange object as \vcn may not be ideal.

\subsection{WISE}
\label{subsec:wise}
The Wide-field Infrared Survey Explorer (WISE) observed \vcn on 
February 20th and 27th of 2010, using about 30 individual exposures 
for their total photometric measurement \citep{WISE,wise_vizier}. 
\vcn was observed as a strong point source in all four of the WISE 
photometric bands $W1$, $W2$, $W3$, and $W4$. 
More than 15\% of the pixels were determined to be saturated in the 
$W1$ and $W2$ bands\citep{wise_vizier}, thus we quote the measured 
photometry as lower bounds of the true flux at these wavelengths.

Once the cryogenic cooling on the WISE satellite became non-operational, 
observations could no longer be performed in the $W3$ and $W4$ bands, 
but the mission continued to observe the infrared sky in the $W1$ and 
$W2$ bands as the NEOWISE mission \citep{neowise}. While the NEOWISE 
mission observed \vcn on several occasions, it was highly saturated in 
all exposures, having derived mean magnitudes of $W1=3.22$ and $W2=2.19$ mag,
which is much brighter than the objects that NEOWISE was designed to measure \citep{neowise}. Regardless, we provide a further analysis 
of the lightcurve created by NEOWISE in Section \ref{subsec:neowise_lc}.

\subsection{SkyMapper}
\label{subsec:skymapper}
The SkyMapper Survey is a photometric survey of the southern sky 
performed on the 1.35-meter telescope at the Siding Spring Observatory
in Australia. The survey uses six photometric bands, which are $u$, $v$, $g$, 
$r$, $i$, and $z$, and they had their first data release in January 
of 2018 \citep{SMDR1}. \vcn was not included in the stellar catalog 
of SkyMapper DR1 despite being observed in all six bands on three separate 
nights in March of 2014. 

To make use of these data, we downloaded the images from the SkyMapper 
website\footnote{\url{http://skymapper.anu.edu.au/image-cutout/}}
and derived magnitudes based on reference to a nearby star which is 
reported in the stellar catalog of SkyMapper with magnitudes in 
all six bands. The reference star we used has associated SkyMapper ID
285891100 and \textit{Gaia} ID 5199012694392465792.  
We then derived fluxes for \vcn in the SkyMapper bands 
by using the AB zero-point magnitudes reported in \cite{SMDR1}. \vcn 
was saturated in all SkyMapper exposures in the $g$, $r$, 
$i$, and $z$ bands, so we only report measurements for the $u$ and $v$ 
bands. We derive $u = 9.08\pm 0.20$ and $v=8.59 \pm 0.17$ mag (in AB system) 
for \vcn.

\subsection{ASAS-SN}
\label{subsec:asassn}
The All Sky Automated Survey for Supernovae (ASAS-SN) is a program aimed 
at searching for and monitoring bright supernovae in the 
variable sky \citep{asassn1}.  The project currently consists of 24 
telescopes which monitor a large fraction of the night sky for variable 
sources down to magnitudes $V<18$ 
\footnote{\url{http://www.astronomy.ohio-state.edu/asassn/index.shtml}}.
\vcn was observed by ASAS-SN from December 28$^{\rm th}$, 2016 to February 
9$^{\rm th}$, 2018 in the $V$ band and was placed in the ASAS-SN variable 
star catalogue as a Gamma Cassiopeia (GCAS) variable star 
\citep{asassn_variables}. Based on communication with the ASAS-SN 
team, this classification is probably not correct, in part due to the 
fact that \vcn is quite bright for the range of magnitudes that ASAS-SN 
observes \citep{Kochanek17}.

It is clear from the ASAS-SN light curve, shown on the right side of the 
top panel of Figure~\ref{fig:summary_plot}, that \vcn remained at an almost 
constant magnitude of $V \sim 8$ mag for a significant portion of the ASAS-SN 
monitoring before beginning to dim by September of 2017. One may be 
worried that the constant magnitude region is simply due to saturation 
of the ASAS-SN observations. Again, based on communication with the ASAS-SN
team, along with the fact that this measured magnitude is consistent with 
the $G$ band magnitude measured by \textit{Gaia}, this flat portion being 
due to saturation seems unlikely, but is still plausible (private 
communication, Tharindu Jayasinghe, ASAS-SN Team). We further 
analyze these data in Section~\ref{subsec:asassn_analysis}.

\subsection{Dark Energy Camera}
\label{subsec:decam}

\vcn was observed by the Dark Energy Camera (DECam), an instrument created as
part of the Dark Energy Survey (DES) \citep{DES1} and mounted on the Blanco 
4-meter telescope at Cerro Tololo Inter-American Observatory (CTIO) in Chile. 
The observation was done as part of a search for substructure around the 
Magellanic clouds
\footnote{\url{https://www.noao.edu/noaoprop/abstract.mpl?2016A-0366}}. 
One exposure in the DES $r$-band, taken in June of 2016 is shown in Figure 
\ref{fig:image_comp}. \vcn is clearly saturated on that image, so we did not 
derive photometric measurements.

\subsection{TESS}
\label{subsec:tess}

The Transiting Exoplanet Survey Satellite (TESS) is a photometric 
survey satellite launched on April 18th of 2018 whose main scientific 
purpose is the discovery of exoplanets \citep{TESS_Ricker}. TESS observes 
single patches of the sky in 27.4 day segments, providing photometric 
measurements at a cadence of about 30 minutes. \vcn was observed in TESS 
Sector 11 from April 22nd to May 20th of 2019. 

\begin{figure*}
    \centering
    \includegraphics{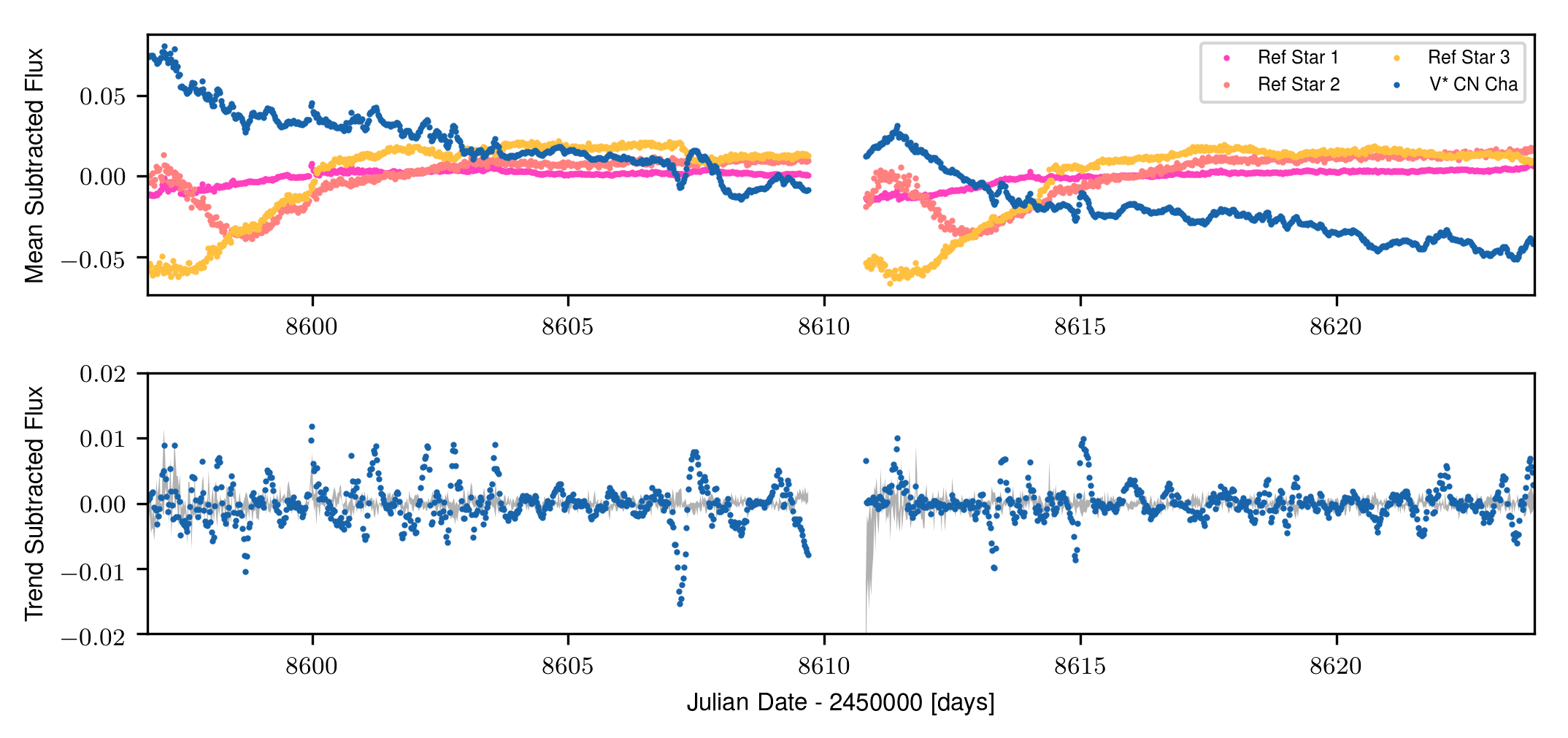}
    \caption{We show the lightcurve inferred from TESS photometry produced as 
    described in Section~\ref{subsec:tess}.
    \textit{Top Panel}: The mean-subtracted flux variations (in units of the 
    logarithm of electron counts on the CCD detector) of \vcn (blue) 
    are compared against three comparison stars (pink and yellow). While all 
    four sources show variations over time, the behavior of \vcn is distinct 
    from that of the reference stars, whose variations are likely due to TESS 
    systematics. It is also clear that \vcn is dimming over the course of the 
    TESS observation, this dimming is still significant over periods where 
    the reference stars are relatively constant in flux. 
    \textit{Bottom Panel}: 
    We show the flux of \vcn (blue) with a running 
    mean on the scale of 10 hours subtracted out so as to make the 
    short-period variations more visible. We additionally show the 1-$\sigma$ 
    variations in trend-corrected flux from the `ensemble' of the three 
    reference stars as the grey shaded region. It is clear that \vcn has 
    significant variation on the timescale of hours.
    }
    \label{fig:tess_lc}
\end{figure*}

To determine the lightcurve (LC) of the star, we have extracted the 100x100 
pixels cutouts from the calibrated full-frame images. The star has $\sim$ 
1200 observations.  Since \vcn has a nearby bright star  
$\sim$ 30\arcsec (1.4 pixels) away, we decided to do a simultaneous PSF fit of the 
16$\times$ 16 pixel area around the star throughout all the epochs, while 
including 3 nearby stars in the model, and allowing for small rotation and 
translation between individual images. The PSF model was a linear combination 
of two Gaussians oversampled at five times the pixel size of the image, that was 
allowed to vary from frame to frame. From this modeling we extract the lightcurve (LC)
of \vcn and of nearby stars. While the proper detrending and 
calibration of the LCs is beyond the scope of this paper, the extracted 
LCs of other stars in the field of \vcn are quite constant with a 
typical median absolute deviation (MAD) of the LC of $<0.5\%$,  the MAD 
of the \vcn LC is 2\% and it shows an almost linear decline in flux 
by about 10\% throughout the time of the TESS observations (27 days). 
That dimming corresponds to roughly a rate of 1.4 mag$\,$yr$^{-1}$ 
(see Figure~\ref{fig:tess_lc}), which is consistent with the dimming starting 
at approximately Julian date 2458000 in the ASAS-SN lightcurve (see Figure 
\ref{fig:asassn_decay}).


\section{Du Pont Echelle Spectroscopy}
\label{sec:our_obs}

The \textit{Gaia} observations outlined above motivated our spectroscopic 
follow up with the Echelle Spectrograph on the 2.5-meter Ir\'en\'ee du Pont 
telescope at Las Campanas Observatory.\footnote{\url{www.lco.cl/telescopes-information/irenee-du-pont}}
This instrument provides coverage in the optical 
from $4000 -9000\,$\AA $\,$ at $R\sim40,000$. Our observations consisted 
of six exposures over 10 minutes from UTC 3:54 to UTC 4:08 on March 12th, 
2019 (Julian date 2458554.672). We performed two 90 second exposures, 
three 60 second exposures and one 10 second exposure. The data was reduced 
using the Du Pont Echelle (DPE) 
reduction pipeline (publicly available online\footnote{\url{code.obs.carnegiescience.edu/dpe-pipeline}}). 
This reduction package makes use of the Carnegie Python Distribution (CarPy) which is also made available on the 
Carnegie Observatories Software Repository\footnote{\url{code.obs.carnegiescience.edu/carnegie-python-distribution}}
\citep{Carpy1,Carpy2}. 

\begin{figure*}
    \centering
    \includegraphics{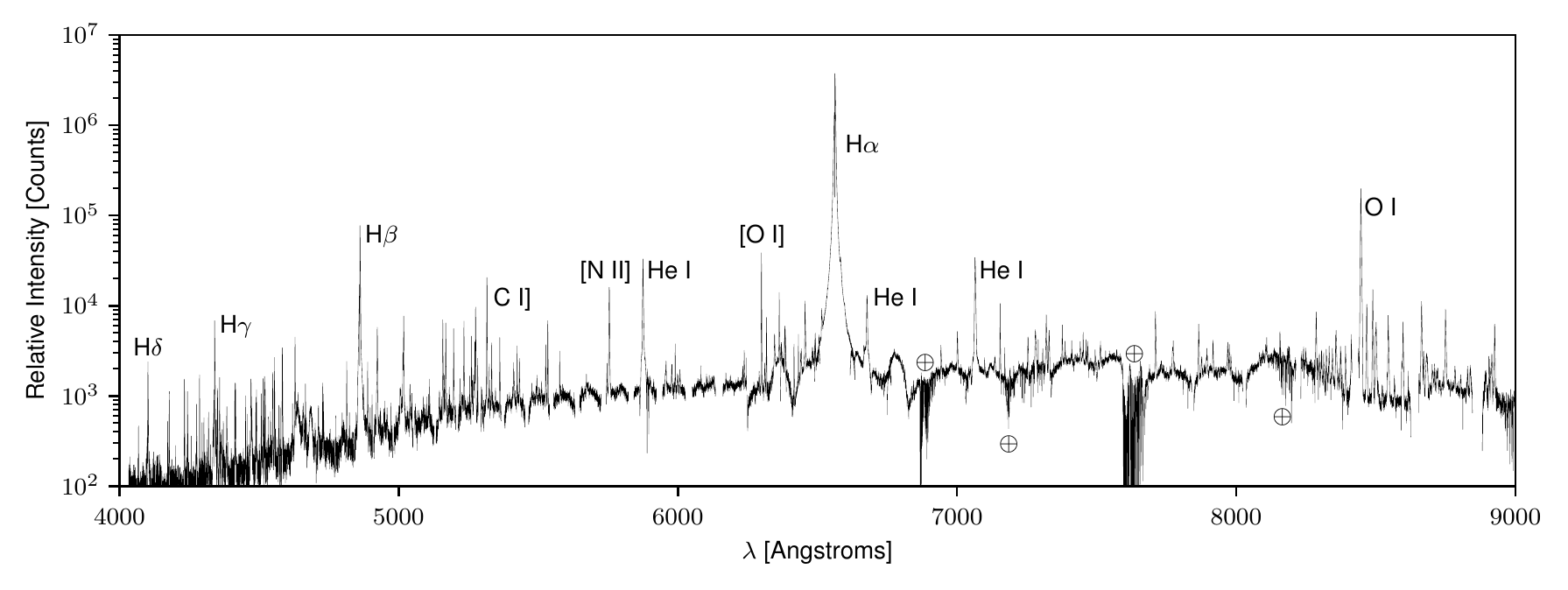}
    \caption{We show the optical spectrum of \vcn observed using the 
    Du Pont Echelle (DPE) instrument on the Ir\'en\'ee du Pont 2.5-meter 
    telescope at Las Campanas Observatory on March 12th, 2019. This spectrum 
    has been reduced 
    using the CarPy software and the reduced spectrum has been divided by 
    a normalized flat in order to account for the changing sensitivity as a 
    function of wavelength. The vertical axis is then given at an arbitrary 
    normalization which can roughly be interpreted as electron counts. We 
    tentatively identify several of the emission lines, including several 
    Helium, Nitrogen, and Oxygen lines and three of the Hydrogen Balmer lines. 
    Regions of significant telluric absorption are indicated with the $\oplus$ 
    sign.}
    \label{fig:dpe_spec}
\end{figure*}

The composite spectrum produced using all six exposures is shown in Figure 
\ref{fig:dpe_spec}. It is clear that the H$\alpha$ emission is quite broad and 
high equivalent width, making up the majority of the star's luminosity in the 
optical. While there are several other emission lines, the
continuum flux is only detected at a signal-to-noise greater than 10 redward 
of approximately $\lambda \sim 4800$\AA. Among the more obvious emission lines 
are the H$\beta$ and H$\gamma$ lines, which stand out clearly. We also 
tentatively identify several of the other prominent emission lines with 
Oxygen ([O I] 6300 \& 6360 along with O I 8446 which is the second strongest 
line), Helium (He I 5875, 6678, and 7065), Nitrogen ([N II] 5756 \& 6585), 
and Carbon (C I] 5317) and label these in \autoref{fig:dpe_spec}.

\section{Further Data Analysis}
\label{sec:deduction}

Here we present some calculations performed on the archival data 
and Du Pont spectrum that go beyond simply presenting the data as 
provided by the various survey teams. We title our subsections by 
the key physical take-aways.

\subsection{Mira Past}
\label{subsec:mira_past}

It seems safe to conclude from several points of observation in the 
archival data laid out above that \vcn was a Mira long-period variable
star in its recent past. This is supported by its ASAS lightcurve shown 
at the left side of the top panel of Figure~\ref{fig:summary_plot} and 
in detail in Figure~\ref{fig:asas_light}. This observation is further 
supported by its archival identification as a Mira variable when it 
was first identified \citep{Hoffmeister63}, its identification as 
a likely variable by IRAS \citep{IRAS}, the NIR colors e.g. in 2MASS 
$J-K=1.69$ mag \citep{DENIS,2MASS}, and the RAVE spectrum gathered in 
early 2010 and discussed briefly in Section~\ref{subsec:RAVE} 
\citep{RAVEDR4}.

\subsection{Slowly Decaying Outburst}
\label{subsec:asassn_analysis}

\begin{figure}
    \centering
    \includegraphics{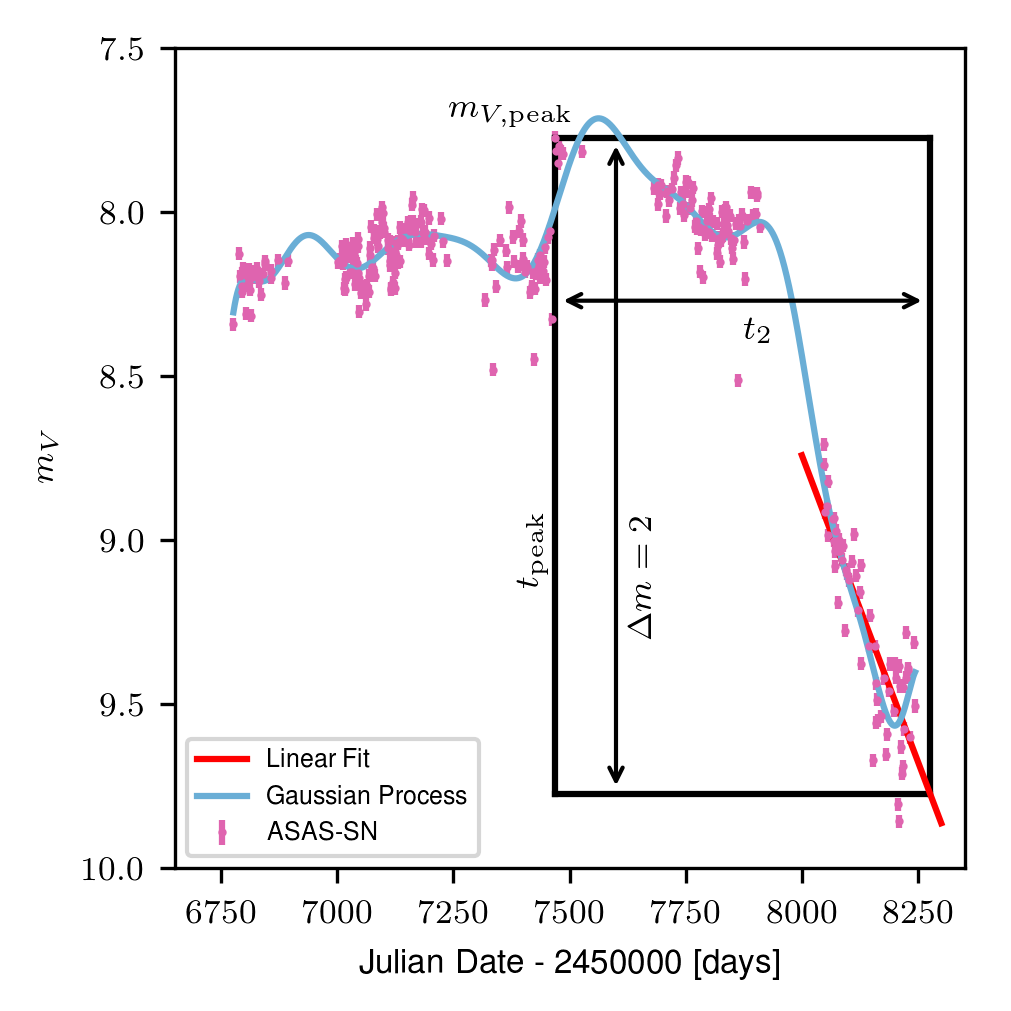}
    \caption{We present an in-depth look at the ASAS-SN lightcurve (shown in 
    purple) for \vcn$\!$. We indicate the peak apparent magnitude (top of black 
    box) time at which the peak magnitude is reached (left side of box), two 
    magnitudes above the peak magnitude (bottom of box) and the time at which 
    this occurs (right side of box). To find the time at which the 
    lightcurve has decayed by two magnitudes we fit a linear decay to the ASAS-SN 
    lightcurve points taken after Julian Day 2458000 and find where this fit 
    (shown in red) crosses 2 mag above the peak apparent magnitude. We 
    additionally we show a Gaussian Process fit to the data in light blue.}
    \label{fig:asassn_decay}
\end{figure}

Here we perform a more detailed analysis of the ASAS-SN lightcurve that was 
presented in Subsection \ref{subsec:asassn}. In particular we want to note 
the peak absolute magnitude and decay time, which are observables typically 
used to compare novae events to one another and are usually quantified using 
the extinction corrected, peak absolute magnitude in the $V$ band 
($M^0_{V,{\rm peak}}$ mag) and the time it takes for the event to decay by two 
magnitudes from this peak ($t_2$). Our analysis is illustrated in 
Figure~\ref{fig:asassn_decay}.

We measure $t_2$ using a linear fit in magnitude space to the lightcurve for 
Julian Date greater than $2458000$ days (the linear decay region). We then 
find the point at which this linear fit crosses two magnitudes decay from the 
peak observed apparent magnitude and define $t_2$ as the difference between 
that crossing time and the time that the peak magnitude was reached. We find
$t_2 = 807\,$days.

We find $M^0_{V,{\rm peak}}$
using the peak apparent magnitude observed in the ASAS-SN lightcurve 
($m_{V,{\rm peak}}=7.773$). We then get the peak absolute magnitude using 
the \textit{Gaia} distance 
\begin{equation}
    M_{V,{\rm peak}}= m_{V,{\rm peak}} -5\log_{10}\left(\frac{d}{10\,{\rm pc}}\right) = -4.74
\end{equation}
and correcting for extinction using $E(B-V)=0.17$ mag
from \cite{SFD_dust} and $R_V = 3.1$ this gives us
\begin{equation}
    \label{eq:peak_asassn_mv}
    M^0_{V,{\rm peak}} = M_{V,{\rm peak}} - R_V E(B-V)  = -5.27 \, .
\end{equation}

Finally, in order to get an estimate of the total amount of energy emitted 
during this outburst event, we calculate the total energy in the $V$ band 
emitted over the course of the ASAS-SN observations. We do this by fitting 
a Gaussian Process (GP) regression, available through the \texttt{scikit learn} 
python package, to the extinction-corrected ($E(B-V)=0.17$) ASAS-SN 
lightcurve \citep{scikit-learn}. This GP fit is 
shown (in observed magnitude space) in Figure~\ref{fig:asassn_decay}. We then 
integrate the fit GP over the full course of the ASAS-SN observation. Using 
this integral in combination with the \textit{Gaia} distance and an effective 
wavelength for the $V$ band of $\lambda_V = 5450$\AA \, we derive the 
total energy in the $V$ band of $E_{V} = 2.8 \times 10^{45}\,$ergs.

\subsection{Continued Bright Emission in MIR}
\label{subsec:neowise_lc}

Though the observations made by the NEOWISE mission of \vcn were saturated 
in all observations, we thought it useful to study the lightcurve if only 
to get a lower bound on the total amount of energy emitted in the $W1$ and 
$W2$ bands over the course of the NEOWISE observations. We restricted our 
analysis to photometric points from the NEOWISE catalog that were within 
3.6" ($0.001^{\circ}$) from \vcn and included the original WISE observations. 
The NEOWISE data are grouped, with several observations occurring over  a few days semi-anually (roughly February and August 2014-2018). As 
these data are quite variable, we average all data points within each group to 
create a smoother overall lightcurve. We then integrate this lightcurve in a 
piecewise linear manner to get a total flux emitted over the observed period.
We find $4.1 \times 10^{45}\,$ergs and $5.4\times 10^{45}\,$ergs as lower 
bounds on the energy emitted in the $W1$ and $W2$ bands respectively.
Additionally, there is no indication that the dimming that is seen in the 
$V$ band is accompanied by dimming in the NEOWISE bands. If any effect 
is observed, it appears to become brighter in $W1$ and $W2$. This is hard to 
interpret given that the images are certainly in the non-linear regime.

\subsection{Balmer Lines: Massive Outflow \& Self-Absorption}
\label{subsec:balmer_lines}

It is clear from the spectrum that the star has an abnormally large Balmer 
decrement. In this subsection we will explore these lines in some detail, 
we first give quantitative measurements of the Balmer decrement and then 
use these numbers to infer the extinction needed to explain these decrements 
solely through dust absorption. We give phenomenological fits to the lines 
and the physical interpretation of these fits. Finally, we also calculate 
the equivalent widths of the lines.

Integrating the flux in each line and comparing we get 
H$\alpha$/H$\beta = 90.8$ (as opposed to the expected value of 3) and 
H$\gamma$/H$\beta \approx 0.08$ (typical value 0.45) \citep{draine_ism}.
We note that the H$\alpha$ emission was saturated in all exposures except for 
the 10 second exposure, so we use this exposure alone 
to measure the H$\alpha$/H$\beta$ ratio. 

This abnormal Balmer decrement would usually be indicative of 
extensive reddening, however, as is noted in \cite{Williams17_BalmerRef}, 
nebulae commonly exhibit strong Balmer decrements which can be due in large 
part to self-absorption in the higher order lines. We will come back to this
in Section~\ref{subsec:nii_ratio}. As a nebula is only one 
of the explanations we explore below, we will nonetheless calculate the 
extinction that would be needed to explain the Balmer decrement that we see 
if there were no self-absorption. Estimating the extinction between 
the H$\alpha$ and H$\beta$ lines as
\begin{equation}
    \label{eq:balmer_dec}
    {\rm E}(\beta - \alpha) = 2.5 \log_{10} 
    \left( \frac{1}{3}
    \frac{F_{{\rm H}\alpha,{\rm obs}}}{F_{{\rm H}\beta,{\rm obs}}}\right) \, ,
\end{equation}
gives E$(\beta - \alpha) = 3.70$ mag. Similarly, we 
estimate E$(\gamma - \beta) = 1.86$ mag. Using the $R_V=3.1$ 
extinction curve from \cite{WD_dust}, we can infer that this extinction 
would imply a visual extinction of $E(B-V)_{\rm Ba} = 3.07$ mag, where 
we have used the subscript Ba to indicate that this extinction measurement 
comes from the Balmer decrement. Using this extinction measurement to correct 
the absolute \textit{Gaia} $G$ band magnitude calculated in section 
\ref{subsec:gaia} would give an intrinsic magnitude of $G_0 = -14.4\,$mag.

Even in the case that $E(B-V)_{\rm Ba}$ is an accurate estimate of the 
true extinction, the above correction is not necessarily accurate given 
the abnormal distribution of light in the $G$ band, as indicated by the 
Du Pont spectrum in Figure~\ref{fig:dpe_spec}. However, this effect alone 
could not change the estimate given above enough to not make \vcn 
intrinsically \textit{very} bright \textit{if} $E(B-V)_{\rm Ba}$ were an 
accurate estimate of the extinction. For this reason it is reasonable 
to conclude that either $E(B-V)_{\rm Ba}$ is not an accurate measure of the 
extinction \textit{or} \vcn is intrinsically as bright as a galaxy. While 
there have been transients with similar lifetimes and peak magnitudes as this,
such as those found in the SPIRITS survey \cite{KasliwalSPIRITS}, we 
determine the former to be the more likely scenario.

\begin{figure*}
    \centering
    \includegraphics{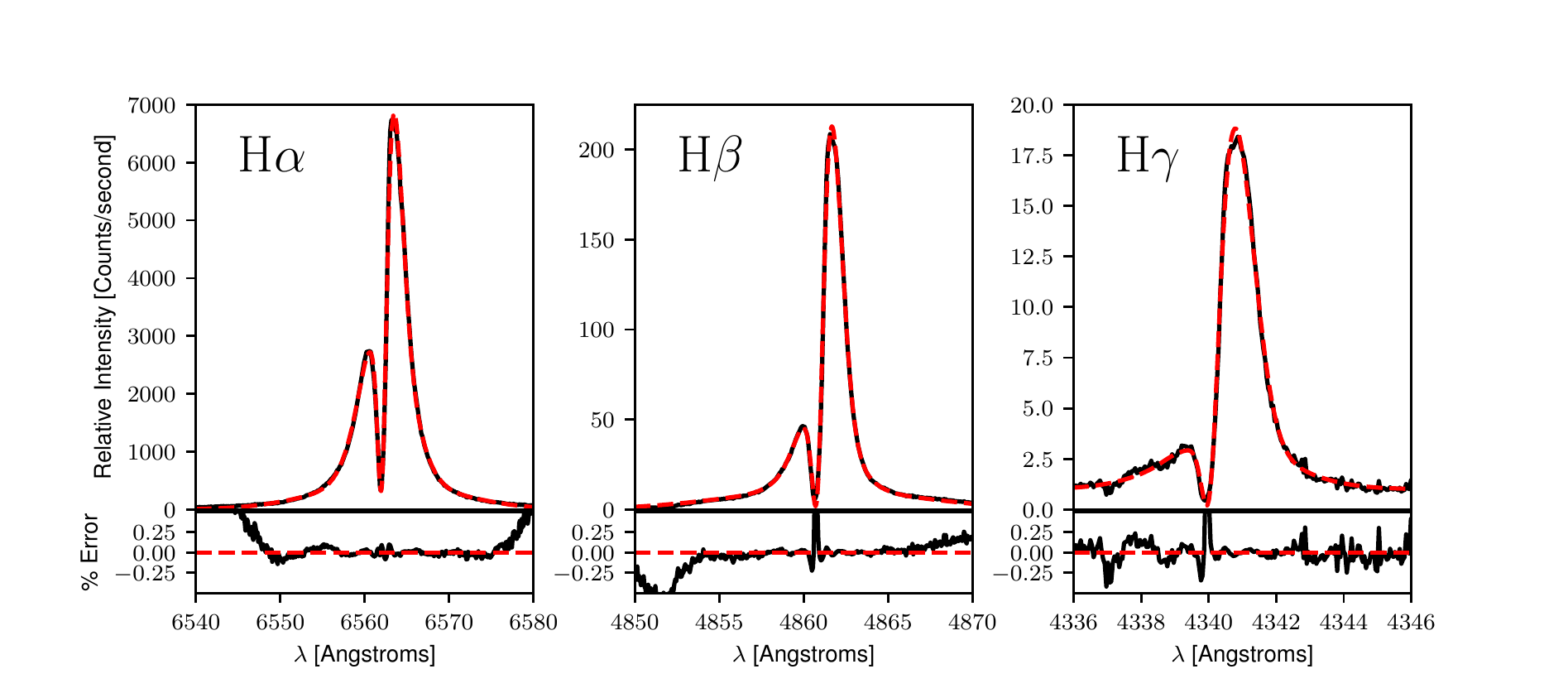}
    \caption{A zoom-in of the Balmer lines seen clearly in the spectrum 
    shown in Figure \ref{fig:dpe_spec}. The data is shown in black while 
    our five component Gaussian fit to the lines is shown as a red, dashed 
    line. From left to right we show the H$\alpha$, H$\beta$, and H$\gamma$ 
    lines. All lines are shown on the same scale, though the H$\alpha$ line 
    is reduced purely from a 10 second exposure, as it was saturated in all 
    60 and 90 second exposures. The bottom panels show the residuals 
    of the data relative to the five component Gaussian fit in terms of percent 
    error.  All lines are clearly quite broad and exhibit blue-shifted 
    absorption, P-Cygni like, profiles.}
    \label{fig:balmer_lines}
\end{figure*}

In Figure~\ref{fig:balmer_lines} we show a zoom-in of the Balmer emission
clearly visible in Figure~\ref{fig:dpe_spec}. Note that while the lines 
are all displayed on the same scale, the H$\alpha$ line is reduced from the 
10 sec exposure alone (since it is saturated in all other exposures) while 
the other lines are simply cutouts of Figure~\ref{fig:dpe_spec}. It is clear 
from inspecting the Balmer lines that they all exhibit strong blue-shifted 
absorption in P-Cygni-like profiles. These profiles are typically indicative 
of a massive outflow of gas from the star. Each line is also very 
broad with widths on the order of $120\,{\rm km}\,{\rm s}^{-1}$ (see below).

Attempts to fit the lines with either single Gaussian profiles or single Voigt 
profiles for the emission and absorption (over the wavelength regions shown in 
Figure~\ref{fig:balmer_lines}) were unsuccessful. In an attempt to interpret 
these profiles shapes, we fit a five component Gaussian to each line. The model 
is defined as
\begin{equation}
    F_{\rm mod} (\nu) = F_{\rm emit}(\nu)
    - F_{\rm abs} (\nu) \, ,
\end{equation}
where $F_{\rm mod}(\nu)$ is the model prediction as a function of frequency,
$F_{\rm emit}(\nu)$ is the component dedicated to capturing the emission profile
and $F_{\rm abs}(\nu)$ is the component meant to capture the absorption profile.
We devote three Gaussians to the emission profile ($F_{\rm emit}(\nu)$) and two
Gaussians to the absorption profile ($F_{\rm abs}(\nu)$). Within the emission or 
absorption profiles respectively all constituent Gaussians are enforced to have 
the same mean frequency, while all Gaussians are allowed to have different 
amplitudes and widths. The parameters of our model are then the central frequency 
of the emission profile ($\nu_e$), the central frequency of the absorption 
profile ($\nu_a$), the amplitudes of the Gaussians for the emission profile 
($A_{e,1}$,$A_{e,2}$,$A_{e,3}$), the widths of the Gaussians for the 
emission profile ($\sigma_{e,1}$,$\sigma_{e,2}$,$\sigma_{e,3}$), the amplitudes 
of the Gaussians for the absorption profile ($A_{a,4}$,$A_{a,5}$), and the 
widths of the Gaussians for the absorption profile ($\sigma_{a,4}$,$\sigma_{a,5}$).
This leaves us with 12 free parameters.

The physically relevant parameters of these fits are given in Table 
\ref{tab:balmer}.  The radial velocity shifts RV$_{\rm emit/abs}$ are derived 
from the central frequencies of the fit profiles $\nu_{a/e}$ and the velocity 
Full-Width at Half-Maximum (FWHM) of the lines is derived by taking the 
frequency difference of the intersection of the profiles with their 
half-maximums. The dimensionless equivalent widths $W$, defined as
\begin{equation}
    W = \int \frac{\diff \nu}{\nu_0} 
    \left(1 - \frac{F_{\nu, {\rm obs}}}{F_{\nu, 0}} \right) \, ,
\end{equation}
where $\nu_0$ is taken to be the rest frame wavelength of the particular 
line in air, $F_{\nu, {\rm obs}}$ is the observed flux, and $F_{\nu, 0}$ is the 
background flux that would have been observed if the line were not present. 
We assume $F_{\nu, 0}$ to be a constant and go about estimating it by taking 
the two Echelle orders on either side of the line of interest, sigma-clipping 
them to remove smaller lines that may be present, and taking the mean of the 
resulting set of data. This method may not be entirely accurate for the 
H$\gamma$ line, where the continuum is not detected at high signal-to-noise.
Finally, the absorption-to-emission ratio given in Table \ref{tab:balmer} 
is estimated by integrating both the fitted emission and absorption profiles 
over the same wavelength regions used for the equivalent width estimates and 
taking the ratio of these integrals. We find the absorption to be between a 
half and a third of the emission across all lines.

\begin{table*}[]
    \centering
    \begin{tabular}{c||c|c|c|c|c|c} 
        Line & RV$_{\rm emit}$ [km/s] & RV$_{\rm abs}$ [km/s] &
        $(\Delta v)_{\rm FWHM, emit}$ [km/s] & $(\Delta v)_{\rm FWHM, abs}$ 
        [km/s] & $W$ & Absorption/Emission \\ \hline
        H$\alpha$ & $-14.04 \pm 0.46$ & $-34.51 \pm 0.19$ 
        & 166.74 & 89.03 & -1.81 & 0.44 \\ \hline 
        H$\beta$  & $-9.26  \pm 1.12$ & $-29.56 \pm 0.18$ 
        & 121.34 & 72.65 & -0.12 & 0.48 \\ \hline
        H$\gamma$ & $ 3.19 \pm  2.67$ & $-28.25 \pm 0.28$ 
        & 118.35 & 59.26 & -0.02 & 0.33
    \end{tabular}
    \caption{The physically relevant parameters of our five-component Gaussian 
    fit to the profiles of the Balmer lines shown in Figure \ref{fig:balmer_lines}.
    The first column specifies the line, columns 2 and 3 give the velocity blue-shifts 
    which were fit to each of the lines in emission and absorption, while columns 4 
    and 5 give the Full-Width at Half-Maximum (FWHM) of the profiles for the lines 
    in both emission and absorption. Column 6 gives the equivalent width of the line 
    and column 7 gives the ratio of absorption to emission in the line, both calculated 
    as explained in the text.}
    \label{tab:balmer}
\end{table*}{}

\subsection{N II Ratio Indicates High Density}
\label{subsec:nii_ratio}

We identify the emission lines [NII] 5756 and [NII] 6585, the latter of 
which sits on top of the wings of the H$\alpha$ line. The ratio of these 
emission lines can be used as a temperature diagnostic \citep{draine_ism}. 
After subtracting out the background by fitting polynomials of degree 2 
and 6 to the regions surrounding the [NII] 5756 and [NII] 6585 lines 
respectively, we sum their fluxes and find a ratio of 2.385. If we 
additionally account for the extinction that occurs between these two 
lines based on the $E(B-V)$ from \cite{SFD_dust} the line ratio becomes:
\begin{equation}
    \label{eq:nii_ratio}
    \frac{\rm [NII] 5756}{\rm [NII] 6585} = 2.583 \, .
\end{equation}
While, as just mentioned, this ratio is usually used as a temperature 
diagnostic, at such large values it is more useful as a density diagnostic
\citep{draine_ism}. Assuming a gas temperature of $T=10^4\,$K this ratio is 
indicative of an electron density of $n_e \gtrsim 10^7\,{\rm cm}^{-3}$ 
\citep{draine_ism}. These high densities would be consistent with 
self-absorption in the Hydrogen recombination lines that we posited in the 
last section \citep{Netzer75,DrakeUlrich80}. 

\subsection{Dynamics of a Thick Disk Star}
\label{subsec:dynamics}

Here we review the dynamics of \vcn within the Galaxy that are derived from 
the astrometric measurement from \cite{GAIA_DR2_contents} and a radial 
velocity of $\sim$0 km/s, motivated by our fits to the Balmer lines given in 
Subsection~\ref{subsec:balmer_lines}.  This is laid out in 
Figure~\ref{fig:dynamics}.

Though \vcn is spatially coincident on the 
sky with the Chamaeleon Complex, a star forming molecular cloud complex, 
lying between the Cha I ($\sim2.5^{\circ}$ separation) and Cha III 
($\sim4^{\circ}$ separation) clouds, its distance as derived from 
its \textit{Gaia} parallax ($3.18^{+0.27}_{-0.23}$ kpc) places it 20$\times$ 
farther than the Chamaeleon 
Cloud ($\sim180\,$pc) \citep{Zucker_CloudDistances}. Furthermore, the 
proper motion of \vcn (a quantity that is in general much better 
measured by \textit{Gaia}) is also inconsistent with the proper motions 
of stars known to be associated with the Chamaeleon Complex. To be  
precise, the measured proper motion of \vcn is 
$(\mu_{\alpha*},\mu_{\delta}) = (-5.42,7.02)\, {\rm mas}\, {\rm yr}^{-1}$
while the proper motions of  HD\,97300 
($(\mu_{\alpha*},\mu_{\delta}) = (-21.01,-0.61)\, {\rm mas}\, {\rm yr}^{-1}$)
and HD 97048 
($(\mu_{\alpha*},\mu_{\delta}) = (-22.44,1.31)\, {\rm mas}\, {\rm yr}^{-1}$), 
both YSOs known to be associated with the Cha I cloud, are significantly different 
with the motion of  V* CN Cha \citep{chamaeleonComplexRef,GAIA_DR2_contents}.

In Figure~\ref{fig:dynamics} we compare the past orbit of \vcn with a star known to 
be in the Chamaeleon Cloud Complex (HD 97048) and the orbit of the Sun.
it seems quite clear that \vcn has dynamics that 
are consistent with a thick disk star, since the maximum vertical extent of it's 
orbit is nearly $2\,$kpc above/below the disk plane and it is moving in a prograde 
fashion. More quantitatively, if we relate it's vertical action 
$J_z = 75.3\, {\rm kpc}\, {\rm km}\, {\rm s}^{-1}$ to it's age using the thin disk 
vertical heating study of \cite{Ting19}, the probability that this star is from the 
thin disk is negligible. This demonstrates that the star is most likely from the 
thick disk and therefore likely more than $8\,$Gyrs old.

\subsection{Stochastic Optical Variability}
\label{subsec:tess_power}

We perform a power-spectrum analysis of the \emph{TESS} light curve 
presented in Section~\ref{subsec:tess} and displayed in Figure~\ref{fig:tess_lc}. We apply 
the \textsc{astropy} Lomb-Scargle periodogram implementation \citep{vanderplas18} 
to the light curves from TESS Sectors~11 and~12 considered separately and together. 
Since strong \emph{TESS} systematics occur on the timescale of its orbit (13.5~days), 
these are filtered out at the calibration stage of our analysis and therefore 
longer-timescale astrophysical variability is also removed.

\begin{figure}
    \centering
    \plotone{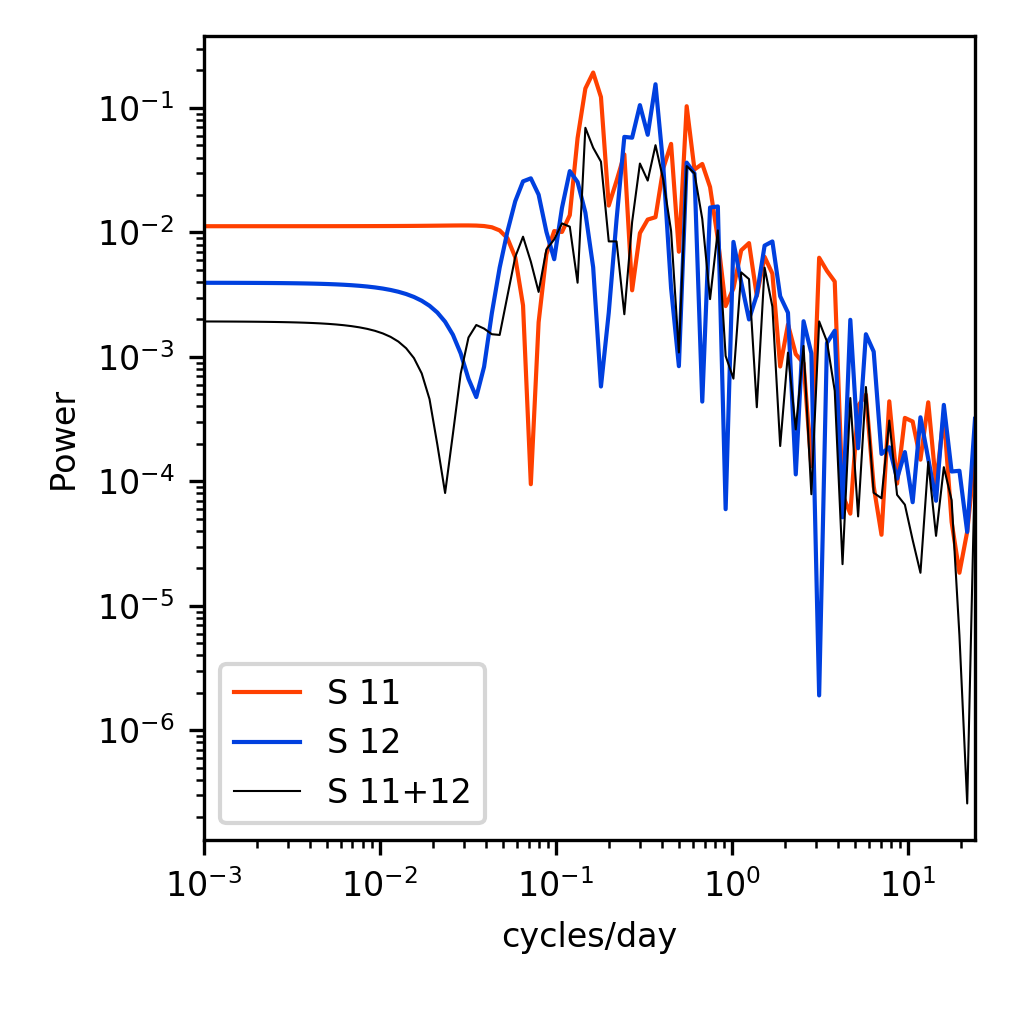}
    \caption{A power spectrum of the lightcurve taken by TESS, which 
    is illustrated in Figure~\ref{fig:tess_pow_spec}. The power spectrum 
    of the data taken in Sector 11 of TESS is shown in red while that for 
    Sector 12 is shown in blue. The joint power spectrum across both 
    sectors is shown in black. All power spectra show red-noise 
    which is indicative of stochastic variability.}
    \label{fig:tess_pow_spec}
\end{figure}

The resulting power spectra are displayed in Figure~\ref{fig:tess_pow_spec} 
and the results are similar for the sectors treated independently. 
The spectrum decays with a red noise or `flicker' profile above about $0.1$ cycles per day;
systematics on this timescale have been filtered out. The 'flicker' is consistent with stochastic 
variability, such as the turbulent motion of a convective photosphere or 
an accretion disk.

\section{Possible Explanations}
\label{sec:theories}

Here we present possible explanations for the observations of this 
object that were outlined in Sections \ref{sec:data}, 
\ref{sec:our_obs}, and \ref{sec:deduction}. We present a simple 
comparison summary of how these theories explain the observations in 
Table \ref{tab:theory_vs_obs}. The theories explained below are ordered 
from least to most likely.

\subsection{Young Stellar Object}
\label{subsec:yso}

Many characteristics of \vcn\ seem to suggest we have caught a Young Stellar 
Object in outburst. In this case the main YSO candidates would be either a 
T Tauri star \citep{MillerTTauriOutburst,AppenzellerTTauri}, a Herbig 
Ae/Be star \citep{HillenbrandHerbigAeBe,WWHerbigAeBeReview}, or an FU Orionis 
type star \citep{HKFuOriRev85,HKFuOriRev96}. Indeed many new young stellar 
objects have been discovered using \textit{Gaia} in combination with archival 
data \citep{HillenbrandG17bpi,HillenbrandG19ajj,HillenbrandPTF14jg}. In 
particular, the time-scale and relatively consistent brightness in the 
lightcurve post-outburst shown in the top panel of 
Figure~\ref{fig:summary_plot} seem to be quite consistent with a period 
of increased accretion that is commonly seen in FU Ori type stars 
\citep{HKFuOriRev96}.

There are two key observations which lead us to believe that a YSO is not 
a sufficient explanation of the data:
\begin{itemize}
    \item[(1)] \vcn appears to have been a Mira type long-period variable 
    star from 1963 to as late as 2010, as laid out in Section \ref{subsec:mira_past}.
    As far as we are aware, there is no YSO object that has photometric 
    variability characteristic of a Mira.
    
    \item[(2)] The dynamics of the star, suggested by \textit{Gaia} proper motions,
    indicate that the star is quite old. This is explained above in 
    Section \ref{subsec:dynamics}.
\end{itemize}

This evidence strongly indicates that \vcn is not some kind of 
YSO. Though it is true that young stars appear everywhere in our galaxy 
\citep{PW1}, they occur with much lower frequency outside of the disk plane 
\citep{splash,thebible}. The fact that, based on \textit{Gaia} dynamics, the 
last time that \vcn crossed the mid-plane of the disk was 52.5 Myrs 
ago is a particularly bad sign for the YSO theory given our understanding of 
star formation mostly occurring in dense molecular clouds which are most common 
in the disk mid-plane \citep{McKeeOstrikerSFRev}.

\begin{figure}
\includegraphics{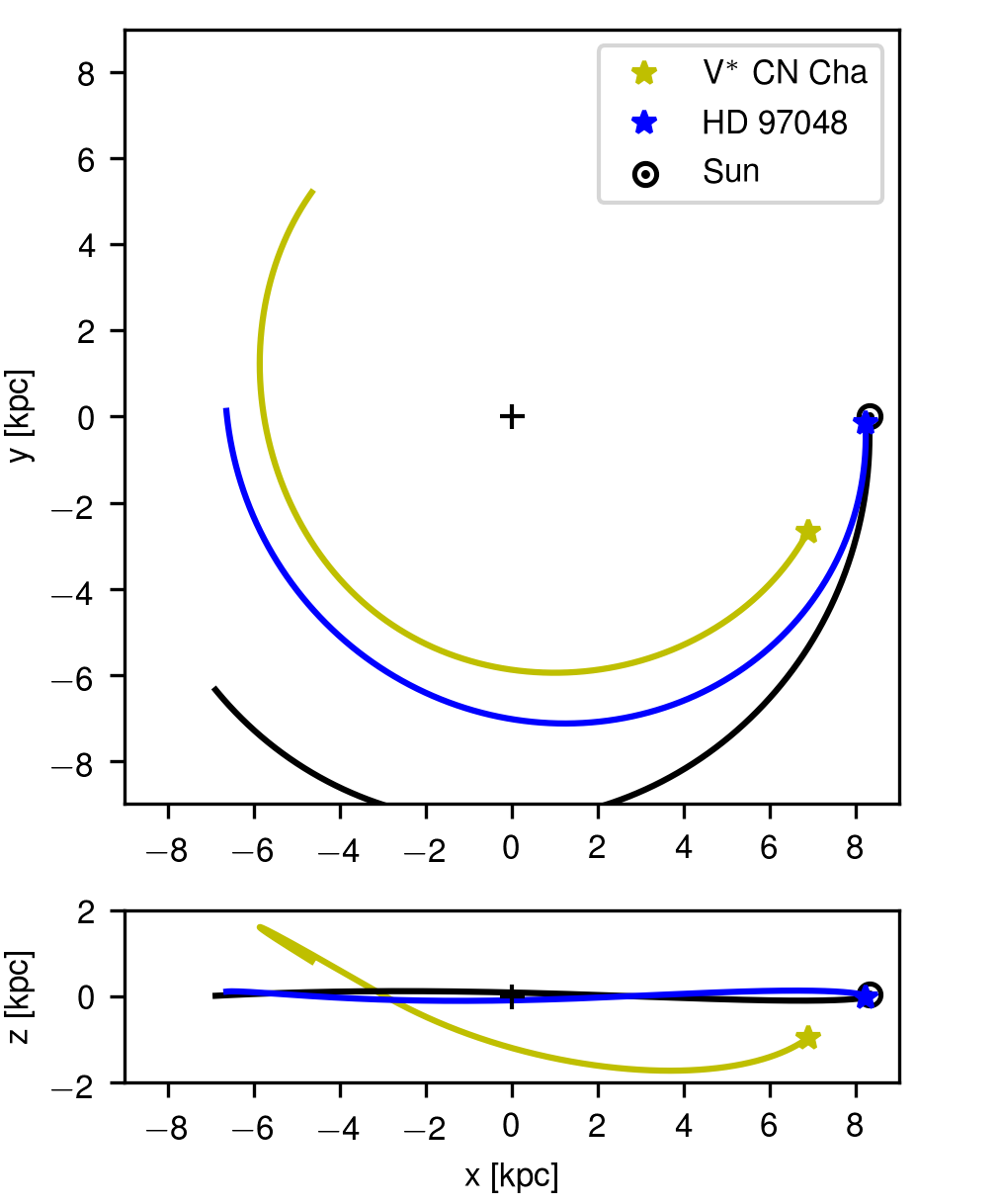}
\caption{We show the galactic orbit of the star \vcn (shown as a yellow star) 
along with the orbits of HD 97048 (a YSO known to be associated with the Chamaeleon 
Complex) in blue and the Sun in black. The top panel looks down up the disk plane, 
face-on while the bottom panel looks edge-on. The Galaxy is centered on the center 
of both panels. the The orbits were integrated 100 Myrs 
back in time using \texttt{galpy} and the 2014 Milky Way Potential from \cite{galpy}.}
\label{fig:dynamics}
\end{figure}

\subsection{Protoplanetary Nebula}
\label{subsection:ppne}

Another possible explanation for the phenomena we have outlined above 
would be the onset of the protoplanetary nebula (PPN) phase at the end of 
an evolved AGB cycle \citep{KwokPPn}. While the observations of a 
massive outflow, combined with \vcn being luminous in the infrared, and the presence 
of strong Hydrogen recombination lines, are consistent with a planetary nebula,
the evolutionary timescales exhibited by \vcn are simply too short to be 
associated with a planetary nebula \citep{Marigo1,Marigo2}. In particular, 
PPN theoretically should not show significant dimming from their peak 
magnitudes until several thousand years after the onset of the PPN phase 
\citep{Marigo1}.

Additionally, our spectra of \vcn exhibit no signs of the [O III]5007\AA 
line that is ubiquitously present in all known planetary nebulae 
\citep{Mendez17}. This would tend to indicate that our system is much 
cooler than the typical proto-planetary nebula.

\subsection{Symbiotic Nova}
\label{subsec:symb_bin}

A symbiotic binary star is a binary star system consisting of an evolved 
star, usually a red giant branch (RGB) star or asymptotic giant branch (AGB) 
star, and a hot, compact companion, usually a white dwarf  
\citep{MP_MiraBins,PM_SBs1}. Binary systems with one component consisting 
of a white dwarf are thought to be the progenitors of classical novae, 
wherein hydrogen-rich material that is accreted from the binary companion 
on to the white dwarf triggers a thermonuclear runaway (TNR) 
\citep{DarnleyHenze19,CNe_review}. Though most classical novae are thought 
to have late-type main sequence star companions \citep{DarnleyHenze19}, 
there are a few nova systems known to contain evolved star companions, and 
they particularly seem to make up the majority of known recurrent novae (RN), 
which have repeating outbursts on human-measurable timescales \citep{RSOph_SNe,Darnley12}.

There are many characteristics that make the observations of \vcn consistent with a nova occurring in a symbiotic binary 
system. The star clearly 
was classified as a Mira-type variable \citep[an evolved AGB star][]{CatelanStars} 
for decades, explained in Section~\ref{subsec:mira_past}. The symbiotic 
binary theory would also explain the UV excess exhibited by the UVOT 
observations, visible in the bottom-left panel of Figure~\ref{fig:summary_plot},
and fit by a 10000$\,$K blackbody. 
The stochastic variability observed in the \emph{TESS} lightcurve and analyzed in 
detail in Section \ref{subsec:tess_power} is well explained by 
an accretion disk around the white dwarf companion. This explanation would 
additionally be consistent with the dynamics of the star outlined in 
Figure~\ref{fig:dynamics}, which indicate it is an old thick-disk star. 

While the timescale of decline for the lightcurve of \vcn as presented by 
the ASAS-SN data is somewhat abnormal for novae in general (typically on 
the order of tens of days), it is not abnormal for symbiotic novae. In 
particular, \vcn has a timescale that is quite similar to the Symbiotic 
Nova in PU Vul, which declined over thousands of days \cite{SyNe10}. The 
timescale for decline is also quite similar to several mid-infrared 
transients which have been observed in extragalactic surveys 
\citep{KasliwalSPIRITS}.

\begin{table*}
    \centering
    \begin{tabular}{c||c|c|c|c|c|c} 
          Theory & Mira & UV  & Outburst & 
          Balmer & Dynamics/ & Balmer \\ 
          & Variability & Excess & Timescale & 
          Decrement &  Age & Line Strength\\ \hline
          YSO & \exmark & \gcheck & \gcheck & \exmark & \exmark & \unsure \\ \hline
          pPN & \gcheck & \unsure & \exmark & \unsure & \gcheck & \unsure \\ \hline
          Symbiotic Nova & \gcheck & \gcheck & \gcheck 
          & \unsure  & \gcheck & \unsure \\ \hline
    \end{tabular}
    \caption{We give a simple comparison between the key observations we 
    have described and the theories we have proposed for explaining them. 
    A \gcheck $\,$ indicates that the theory explains the observation, a \exmark $\,$
    indicates that the theory does not explain the observation, and a \unsure  $\,$
    indicates that it is uncertain whether or not the theory explains the 
    observation.}
    \label{tab:theory_vs_obs}
\end{table*}

\section{Discussion}
\label{sec:discussion}

Of the explanations presented in Section~\ref{sec:theories} we believe that 
a symbiotic novae is most consisted with the wide range of 
observations we have gathered on \vcn. In this section, we briefly 
discuss the implications for the interpretation of \vcn as a symbiotic nova.

One of the key empirical insights that has resulted from the study of 
classical novae is the relationship between their peak magnitudes and 
the time-scale over which they decay, known as the Maximum Magnitude 
Rate of Decline (MMRD) relation \citep{DownesDuerbeck}. This empirical 
relationship shows that brighter novae decay more quickly to their 
pre-nova magnitudes. The mass of the white dwarf in the system is thought to drive the MMRD \citep[e.g.,][]{Livio92},
such that the higher the white dwarf mass is, the higher 
the surface gravity and hence the higher the pressure in the white dwarf 
atmosphere, causing a stronger thermonuclear runaway (TNR) and therefore 
a brighter peak luminosity. At the same time, at higher white dwarf mass 
the less massive the atmosphere needs to be before TNR is triggered, 
causing the nova to decay more quickly, as it has less fuel.

The discovery in recent years of several novae that fall off of the MMRD 
relation has called into question the simple argument outlined above 
\citep{Kasliwal11,Shara17}. This has mainly been driven by the further 
study of galactic recurrent novae \citep{HachisuKato19} and the discovery 
of so-called `faint-fast' novae in M31 and M87. In particular, it has been 
argued that the MMRD should at the very least depend upon some details 
of the donor star as well as the white dwarf's temperature and composition. 

If we take the peak absolute magnitude in the $V$ band of \vcn to be 
$M_{V,{\rm peak}}^0 = -5.27$ mag and its time-scale to decay from this peak by 
two magnitudes to be $t_2 = 807\,$days (calculations outlined in 
Section~\ref{subsec:asassn_analysis}), the outburst we observe would be one 
of the longest decays currently measured for novae.
The most direct comparison, is the Galactic symbiotic nova,
PU Vul, and some infrared-discovered extragalactic transients \citep{SyNe10,KasliwalSPIRITS}. 
Furthermore, \vcn is among the lowest luminosity of any known nova, roughly tying for this distinction 
with novae 39-41 of \cite{Shara16}.

Symbiotic binary systems are also thought to be possible progenitors for Type Ia 
supernovae \citep{RSOph_SNe,Sym_SNeIa}. Recent work has even suggested that mass accretion 
onto the compact companion via so-called Wind Roche-Lobe Overflow (WRLOF) could 
make these binary systems even more likely to create supernovae \citep{MP_WRLOF,WRLOF2}.
In this context, \vcn may be able to provide interesting insights in to the 
viability of this progenitor channel for Type Ia supernovae.



\section{Summary and Follow-Up}
\label{sec:conclusion}

We have presented and organized the various archival data on the stellar 
object \vcn in the constellation of Chamaeleon. We used these observations 
to motivate that the luminosity from \vcn was dominated by a Mira long-period 
variable star from 1963 until 2013, at which point the star seemed to enter an 
outburst phase. After remaining at roughly a constant luminosity for 
about 3 years, the object then entered a dimming phase, during which time 
it dimmed at about 1.4 magnitudes per year.

We then presented the optical spectrum that motivated our initial interest in this system. 
The spectrum is dominated by emission in hydrogen 
recombination lines, predominantly from H$\alpha$. The Balmer series lines  
exhibit blue-shifted absorption, indicative of a massive outflow 
from the object. The spectrum additionally presents emission lines of He I, 
C I], [N II], O I, and [O I], among many others that we did not 
explicitly identify. Many of these lines exhibit complicated profiles.

We estimate the electron density of the environment using the ratio 
of [N II] forbidden lines and we find a density of
$n_e \gtrsim 10^7 \, {\rm cm}^{-3}$. We additionally concluded that the 
extreme Balmer decrement seen in our spectrum was most likely not due to 
extinction but due to self-absorption; this is consistent with the high 
densities indicated by the [N II] ratio. 
We made phenomenological fits to the profiles of the Balmer series lines 
which indicated large velocity widths ($\sim120\,$km/s) and outflow 
velocities from the absorption of $\sim20\,$km/s. We additionally provide 
an analysis of the peak magnitude and decay time of the outburst, indicating a rather long period and low peak-luminosity compared to 
other novae outbursts. Finally, we showed that the dynamics of the 
object, as inferred from the \textit{Gaia} astrometry and a radial 
velocity, indicate an old star, perhaps a 
component of the thick disk (lying $1\,$kpc below the disk mid-plane).

After considering that \vcn may be either a young stellar object or an early 
phase protoplanetary nebula, we used the evidence summarized above to argue 
that the data are best explained by a nova occurring in a symbiotic binary 
system. If this is indeed the case, the \vcn outburst would be among the 
lowest luminosity novae ever documented (tied with novae 39-41 of 
\cite{Shara16}), providing interesting new context to the MMRD relation 
for novae.

More concrete conclusions will be drawn about this object from an in-depth 
analysis of the spectral data that we have gathered, including 
follow-up high-cadence spectra at both high and moderate resolution that 
we gathered in late November/early December of 2019. We reserve detailed 
analyses of this data to future work. 

We will be also able to learn a great deal more about this object's evolution 
over the course of its outburst once the \textit{Gaia} BP/RP spectra (which 
cover almost the entirety of the outburst phase) are released to the community;
this is anticipated with \textit{Gaia} DR3 planned for late 2021. Supplementing 
our current data with observations in the radio and X-ray will be extremely 
informative in discerning between the different scenarios we have proposed here.

As we pointed out in the introduction to this manuscript, there is no single 
database that encapsulates the vast array of human-collected astronomical 
data. We believe that the discovery we have described in this work presents 
a concrete example of the consequences of not having such a resource. Though 
nearly all of the information that made this object interesting was freely 
available to the community, it was scattered between many catalogs, interfaces, 
and repositories. As a result, this object was overlooked and its discovery 
was serendipitous. This paper by no means constitutes the sole such example 
\citep{HillenbrandG17bpi,HillenbrandG19ajj,HillenbrandPTF14jg}.

This problem will only become worse in the near future with large upcoming 
time-domain astronomical surveys such as the Vera C. Rubin Observatory 
(VRO, formerly LSST) and large spectroscopic surveys such as SDSS-V, 
\textit{WEAVE}, and \textit{4MOST} \citep{LSST19,SDSS-V,WEAVE,deJong_4MOST}. 
We believe that the importance of well thought-out and easily 
searchable/accessible data will be paramount in the coming astronomical age, 
and that doing this correctly should be considered a priority for the community.

The desire for a tool that would facilitate the access of data across 
surveys of widely varying structure is certainly not a new concept. In fact, 
this study has made use of several such tools intended for this purpose 
and developed over the past generation, including: the ``Whole Sky Database" 
created by Sergey Koposov and maintained at the University of Cambridge \citep{WSDB},
the SIMBAD database created by the Centre de Donn\'ees astronomique de Strasbourg 
(CDS) \citep{SIMBAD}, the Vizier online database of catalogues \citep{Vizier}, and 
the sky visualizer Aladin \cite{Aladin}. There have been several other great efforts 
made by researchers along these lines including the National Virtual Observatory 
\citep{Virtual_Observatory}, the CDS overall \citep{CDS1,CDS2}, the Mikulski 
Archive for Space Telescopes (MAST) \citep{MAST} along with too many others 
to give a complete list here.

Despite the largely successful efforts of these programs, we believe this 
work demonstrates that there are still regions of discovery which are made less 
accessible by the lack of a perfectly effective cross-survey search tool.
Though it is difficult to find and characterize anomalous objects, these 
objects have the potential to make us fundamentally reconsider our 
understanding of the world around us. We look forward to an age of astronomy in 
which these discoveries can be made commonplace.

\acknowledgments

The authors would like to thank Howard Bond, Matteo Cantiello, and Adam Jermyn 
for pointing us towards symbiotic binary systems as a possible explanation for 
this object and for further useful discussions on the nature of the object. 
We would additionally like to thank Michael Shara for recommending relevant 
literature references with regards to symbiotic novae as novae in general.
We would like to thank Tharindu Jayasinghe and Krzysztof Stanek for consulting 
us on the ASAS-SN observations. We would like to thank Arne Henden and 
Sara Beck for consulting us on the APASS measurements. The authors would 
additionally like to thank Gaspar Bakos, Adam Burrows, Trevor David, 
Bruce Draine, Andy Goulding, Emily Levesque, Erin Kado-Fong, Eliot Quataert, 
Nathan Smith, David Spergel, Joshua Winn, and Vasily Belokurov, Wyn Evans and 
the other members of the Cambridge Streams group for useful conversations and 
references. L.L. would like to thank the Center for Computational 
Astrophysics at the Flatiron Institute in New York City for their 
hospitality while part of this work was completed.

This work was performed in part under contract with the Jet Propulsion 
Laboratory (JPL) funded by NASA through the Sagan Fellowship Program 
executed by the NASA Exoplanet Science Institute. Support for this work 
was provided by NASA through Hubble Fellowship grant \#51386.01 awarded 
to R.L.B. and grant \#51425.001 awarded to Y.S.T. by the Space Telescope 
Science Institute, which is operated by the Association of  Universities 
for Research in Astronomy, Inc., for NASA, under contract NAS 5-26555. SK 
is partially supported by NSF grants AST-1813881, AST-1909584 and 
Heising-Simons foundation grant 2018-1030. 

This publication makes use of data products from the Wide-field 
Infrared Survey Explorer, which is a joint project of the University 
of California, Los Angeles, and the Jet Propulsion Laboratory/California 
Institute of Technology, funded by the National Aeronautics and 
Space Administration.

This work has made use of data from the European Space Agency (ESA) mission
{\it Gaia} (\url{https://www.cosmos.esa.int/gaia}), processed by the {\it Gaia}
Data Processing and Analysis Consortium (DPAC,
\url{https://www.cosmos.esa.int/web/gaia/dpac/consortium}). Funding for the DPAC
has been provided by national institutions, in particular the institutions
participating in the {\it Gaia} Multilateral Agreement.

This research made use of NASA's Astrophysics Data System and the 
SIMBAD database, operated at CDS, Strasbourg, France. Some of the data 
presented in this paper were obtained from the Mikulski Archive for 
Space Telescopes (MAST). STScI is operated by the Association of 
Universities for Research in Astronomy, Inc., under NASA contract 
NAS5-26555. Support for MAST for non-HST data is provided by the 
NASA Office of Space Science via grant NNX13AC07G and by other grants 
and contracts. 

This paper made use of the Whole Sky Database (wsdb) created by Sergey 
Koposov and maintained at the Institute of Astronomy, Cambridge by 
Sergey Koposov, Vasily Belokurov and Wyn Evans with financial support 
from the Science \& Technology Facilities Council (STFC) and the European 
Research Council (ERC).

\appendix 

\section{ASAS Catalogues}
\label{asas_app}

Here we briefly discuss the two separate photometric catalogues that 
we found describing the light curve of \vcn based on the ASAS survey.
The first, which included photometric 
data up October 11th, 2003, was obtained from the online catalogue 
\texttt{Vizier} (catalogue 1)\citep{asas_vizier}. The second 
catalogue is the ``ASAS All Star Catalogue" (catalogue 2) which is 
hosted on servers at the University of Washington \citep{asas_uw_cat}. 
This second catalogue contains photometric information over the full 
range of observations, with magnitudes obtained through measurement 
in four separate apertures. When using the data we simply take the 
error-weighted mean of these four measurements.

\begin{figure}
\includegraphics{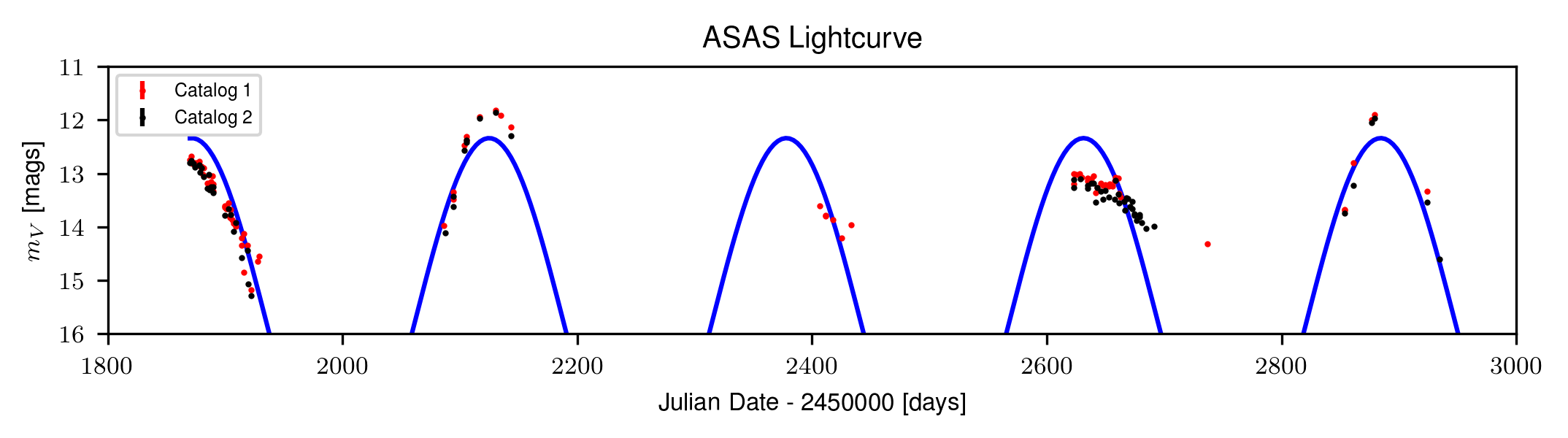}
\caption{We show the ASAS light curve for \vcn in the $V$ band for both 
catalogue 1 (red points) and catalogue 2 (black points) as referred to in 
the text, though only over the period covered by catalogue 1. Both catalogues 
have typical photometric errors of 0.02 magnitudes. To guide the eye we also
include a sine curve with a period of 253.24 days and a mean of 12.34 mags 
(as determined by the original ASAS team, blue curve) we fit for the optimal 
phase shift and amplitude (3.43 mags).}
\label{fig:asas_light}
\end{figure}

When looking at the data in catalogue 2 which was taken during the 
same time interval as is covered in catalogue 1, one would expect 
these data to agree with one another. However, there seem to be 
several data points in catalogue 2 which do not appear in catalogue 
1 and vice versa. Additionally, there are some minor inconsistencies 
in magnitudes between measurements in each catalogue that are meant 
to be measured at the same time. As we weren't able to resolve which 
of these measurements to trust more, we use the data given in both 
catalogues here. A comparison of the two catalogues, over the 
period covered by catalogue 1, is given in Figure \ref{fig:asas_light}.

\bibliographystyle{aasjournal}
\bibliography{bibliography}

\end{document}